\documentclass[aps,prb,twocolumn,superscriptaddress,showpacs,showkeys,floatfix]{revtex4}

\usepackage{graphicx,color}
\usepackage{multirow,slashbox}

\graphicspath{{figs/}}
\begin{document}
\title{A Gaussian Treatment for the Friction Issue of Lennard-Jones Potential in Layered Materials: Application to Friction between Graphene, MoS$_{2}$ and Black Phosphorus}
\author{Jin-Wu Jiang}
    \altaffiliation{Corresponding author: jwjiang5918@hotmail.com}
    \affiliation{Shanghai Institute of Applied Mathematics and Mechanics, Shanghai Key Laboratory of Mechanics in Energy Engineering, Shanghai University, Shanghai 200072, People's Republic of China}
\author{Harold S. Park}
    \affiliation{Department of Mechanical Engineering, Boston University, Boston, Massachusetts 02215, USA}

\date{\today}
\begin{abstract}

The Lennard-Jones potential is widely used to describe the interlayer interactions within layered materials like graphene.  However, it is also widely known that this potential strongly underestimates the frictional properties for layered materials. Here we propose to supplement the Lennard-Jones potential by a Gaussian-type potential, which enables more accurate calculations of the frictional properties of two-dimensional layered materials.  Furthermore, the Gaussian potential is computationally simple as it introduces only one additional potential parameter that is determined by the interlayer shear mode in the layered structure. The resulting Lennard-Jones-Gaussian potential is applied to compute the interlayer cohesive energy and frictional energy for graphene, MoS$_2$, black phosphorus, and their heterostructures.

\end{abstract}

\pacs{63.22.Np, 68.35.Af, 63.22.-m, }
\keywords{Lennard-Jones Potential, Layered Structure, Friction, Shear Mode}
\maketitle
\pagebreak

\section{Introduction}

In 1924, Lennard-Jones published a 12-6 pairwise potential to describe the van der Waals interaction between two atoms, which is now known as the Lennard-Jones (LJ) potential.\cite{JonesJE1924} The LJ potential depends only on the distance ($r$) between two interacting atoms. For a relative displacement $\vec{u}$ between these two atoms, the variation in the distance is $dr=\vec{u}\cdot\hat{e}_{r}$, with $\hat{e}_{r}=\vec{r}/r$, which shows that this potential is able to effectively control the cohesive motion between two atoms. However, the LJ potential cannot describe the frictional motion of two atoms, because a weak relative frictional motion between two atoms does not alter their distance.

This friction issue can be greatly amplified when the LJ potential is applied to the interlayer interaction in quasi-two-dimensional layered materials such as bilayer graphene.\cite{GeimAK2007nm,NetoAHC2011rpp} In these layered structures, the van der Waals interlayer interaction is much weaker than the covalent intra-layer interaction,\cite{JiangJW2008prb} leading to two distinct characteristic types of motion in these layered materials, i.e., the relative cohesive motion and the frictional motion. In bilayer graphene, the LJ potential can describe the cohesive motion accurately, but it is not able to provide an accurate measure of the frictional energy.\cite{LebedevaIV}

There are only two parameters in the LJ potential - one ($\sigma$) is a length parameter determining the interlayer spacing for bilayer graphene, while the other ($\epsilon$) is an energy parameter.  However, bilayer graphene has two independent interlayer motions, i.e., the cohesive motion and the frictional motion. As a result, it is not surprising that the LJ potential cannot describe both the cohesive and frictional motion simultaneously. Several works have shown that this friction issue in the LJ potential for bilayer graphene can be eliminated by introducing seven more potential parameters.\cite{KolmogorovAN2000,KolmogorovAN2005,LebedevaIV}

The aim of the present work is to present a concise supplement for the LJ potential in layered materials while introducing a minimum number of fitting parameter, with the specific goal of accurately capturing the frictional motion.  This would be computational beneficial as it can be readily implemented in most atomistic simulation packages that use the LJ potential.  On the other hand, many advanced properties have been found for the layered materials, which have garnered both academic and industrial attention. For instance, few-layer graphene can serve as an ideal platform for the investigation of some dimensional crossover phenomena.\cite{GhoshS,MakaKF2010pnas,ShangJ2013jrs} It was found that heterostructures like graphene/MoS$_{2}$ can mitigate the less desirable properties of each individual constituent.\cite{BritnellL2013sci,ZanR2013acsn} Hence, it is important to describe the interlayer energy for layered materials more accurately, including the important frictional properties.\cite{VanossiA2013rmp} 

In this paper, we propose to combine the LJ potential with a Gaussian-type potential (LJ-G) to describe the interlayer energy for layered materials. The Gaussian potential introduces only one additional parameter, which is determined by the interlayer shear (C) mode in the layered structure. The LJ-G potential thus has minimum number of potential parameters and can be applied to compute the cohesive energy and frictional energy in graphene, MoS$_2$, black phosphorus (BP), and their heterostructures.  Due to intrinsic lattice mismatch, the frictional energy in all heterostructures is found to be one order lower than the individual constituent.  

\section{Interlayer Potential}

\begin{figure}[tb]
  \begin{center}
    \scalebox{1}[1]{\includegraphics[width=8.0cm]{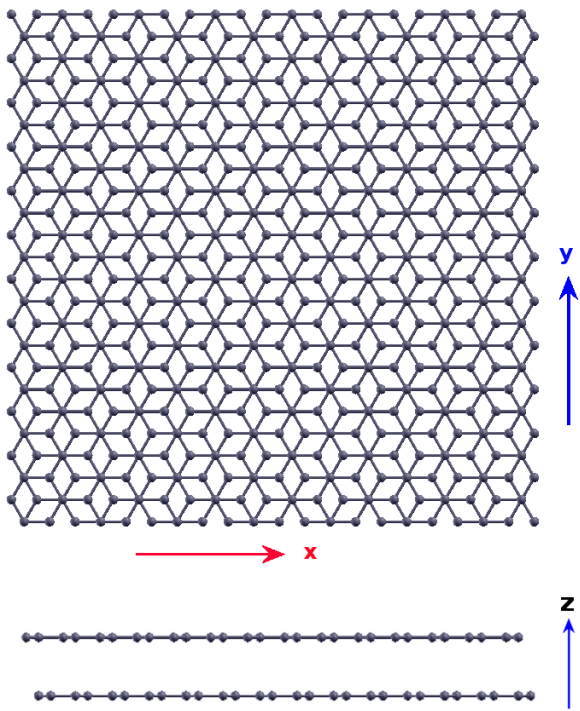}}
  \end{center}
  \caption{(Color online) Top and side views for bilayer graphene of dimension $30\times30$~{\AA}.}
  \label{fig_cfg}
\end{figure}

\begin{figure}[tb]
  \begin{center}
    \scalebox{1.0}[1.0]{\includegraphics[width=8.0cm]{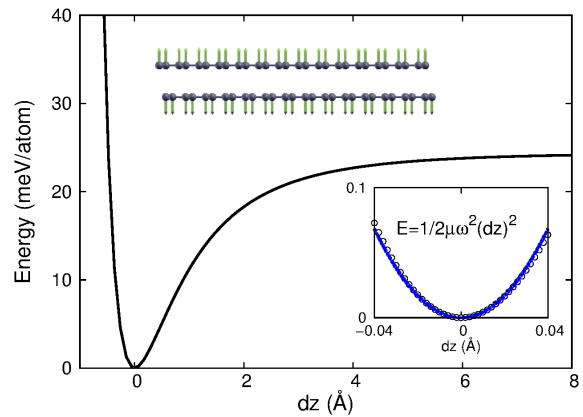}}
  \end{center}
  \caption{(Color online) The energy variation for bilayer graphene with respect to different interlayer spacings. $dz=0$ corresponds to the equilibrium interlayer spacing of 3.365~{\AA}. The interlayer energy is described by the LJ potential with parameter values given below Eq. (\ref{eq_lj}).  The bottom inset shows a zoom in of the curve around the minimum energy point, where the energy variation is fit to a quadratic function $E=\frac{1}{2}\mu\omega^2(dz)^2$. $\mu=m_{\rm gra}/2$ is the effective mass for two vibrating graphene layers, with $m_{\rm gra}$ as the mass of a single-layer graphene. The fitting parameter $\omega=88.4$~{cm$^{-1}$} gives the vibration frequency for the B mode as shown in the top inset.}
  \label{fig_lj_bmode}
\end{figure}

\begin{figure}[tb]
  \begin{center}
    \scalebox{1.0}[1.0]{\includegraphics[width=8.0cm]{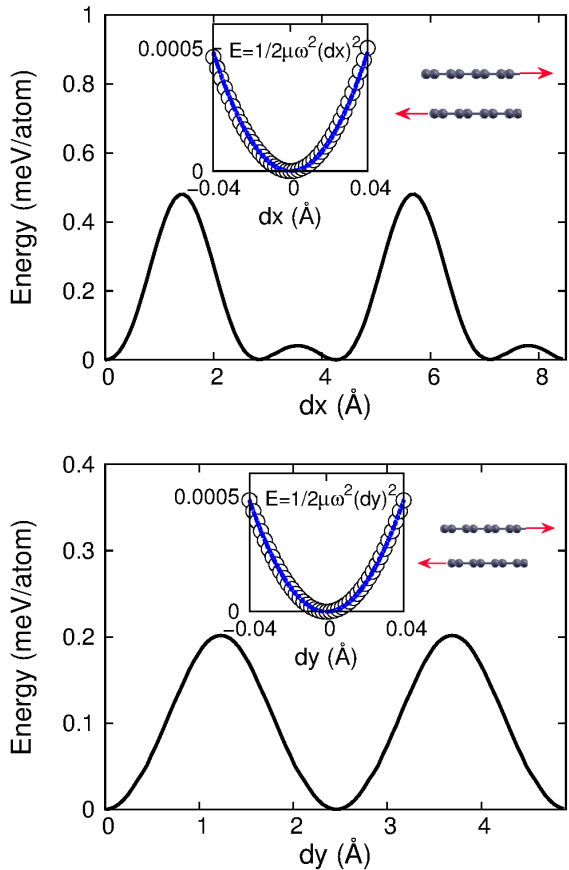}}
  \end{center}
  \caption{(Color online) The energy variations for bilayer graphene as a function of the relative displacement of two layers in the x (top) and y (bottom) directions. The interlayer interaction is described by the LJ potential. Left insets in both panels show the zoom-in of the small displacement regime, where the energy variation is fitted to quadratic functions $E=\frac{1}{2}\mu\omega^2(dx)^2$ and $E=\frac{1}{2}\mu\omega^2(dy)^2$. The fitting parameter $\omega=7.3$~{cm$^{-1}$} gives the vibrational frequency for the C$_{x}$ and C$_{y}$ modes as shown in the right insets. }
  \label{fig_lj_cmode}
\end{figure}

Fig.~\ref{fig_cfg} shows the AB-stacking for bilayer graphene of dimension $30\times 30$~{\AA}. Both top view and side view are shown in the figure. The z-axis is perpendicular to the graphene plane. The x direction is along the armchair direction, while the y-direction is in the zigzag direction.

The following LJ potential is applied to describe the interlayer energy,
\begin{eqnarray}
V_{LJ} = 4\epsilon\left[\left(\frac{\sigma}{r}\right)^{12}-\left(\frac{\sigma}{r}\right)^{6}\right],
\label{eq_lj}
\end{eqnarray}
where $r$ is the distance between two interacting atoms. $\epsilon=2.96$~{meV} and $\sigma=3.382$~{\AA} are potential parameters.  Specifically, the length parameter $\sigma$ is fit to the out-of-plane lattice constant in bulk graphite, while the energy parameter $\epsilon$ is fit to the interlayer breathing (B) mode in bilayer graphene, as will be described in further detail below.

To explore the relationship between $\epsilon$ and B mode, we need to calculate the cohesive energy in the bilayer graphene. The structure is optimized via energy minimization, after which the cohesive energy for bilayer graphene can be computed by evaluating the energy for different separation distances of the individual layers. Fig.~\ref{fig_lj_bmode} shows the cohesive energy for bilayer graphene. The x-axis ($dz$) is the variation in the interlayer spacing with respect to its equilibrium value, so $dz=0$ corresponds to the optimized interlayer spacing.

From lattice dynamical analysis,\cite{BornM} it can be shown that the strength of the relative cohesive motion is related to the frequency of the B phonon mode in bilayer graphene. The vibration morphology of the B mode is shown in the top inset of Fig.~\ref{fig_lj_bmode}. More specifically, for the cohesive energy curve around the minimum energy minimum ($dz=0$), the structure deviates only slightly from its optimized configuration. We can thus consider this small cohesive motion as a linear vibration of the B mode. Hence, we can extract the frequency for the B mode from the cohesive energy as shown in the bottom inset of Fig.~\ref{fig_lj_bmode}, by fitting the cohesive energy to the quadratic function $E=(1/2)\mu\omega^2(dz)^2$. The quantity $\mu=m_{\rm gra}/2$ is the effective mass for two vibrating graphene layers, with $m_{\rm gra}$ as the mass for a single layer of graphene. The fitting parameter $\omega$ yields the B mode's frequency.

\begin{table}
\caption{LJG parameters for bilayer graphene. Numbers in the parentheses are experimental out-of-plane lattice constant\cite{SaitoR} and frequency.\cite{MohrM2007prb} The first row is for the LJ potential ($g=0$), while the second row is the LJ-G potential. Energy parameter is in meV. Length parameter is in \AA.  Frequency is in cm$^{-1}$.}
\label{tab_ljg_gra}
\begin{tabular}{|c|c|c|c|c|c|}
\hline 
$\epsilon$   & $\sigma$  & $g$  & $c$ (6.7)  & $\omega_{B}$ (89.5) & $\omega_{C}$ (37.1)\tabularnewline
\hline 
\hline 
2.96 & 3.382 & 0 & 6.73 & 88.4 & 7.3\tabularnewline
\hline 
2.96 & 3.382 & 94.87 & 6.73 & 88.4 & 37.1\tabularnewline
\hline 
\end{tabular}
\end{table}

We thus fit parameter $\epsilon$ in the LJ potential to the frequency of the B mode in the bilayer graphene. Tab.~\ref{tab_ljg_gra} shows the fitted LJ parameters for bilayer graphene, where the length parameter $\sigma$ in the LJ potential is fit to the out-of-plane lattice constant in graphite. The fitted LJ potential yields $\omega=88.4$~{cm$^{-1}$} for the B mode and the out-of-plane lattice constant $c=6.73$~{\AA}. These results are in good agreement with the experimental results.\cite{MohrM2007prb,SaitoR} It should be noted that the experimental frequency for the B mode in bulk graphite ($\omega_{\rm bulk}$) has been used to extract the frequency of bilayer graphene ($\omega_{\rm bi}$) through $\omega_{\rm bi}=\omega_{\rm bulk}/\sqrt{2}$ according to the linear chain model.\cite{TanPH,ZhaoYY,JiangJW2014BCmode} For instance, experiments found $\omega_{\rm bulk}=126.6$~{cm$^{-1}$} in graphite,\cite{MohrM2007prb} so the frequency of the B mode in bilayer graphene is $\omega_{\rm bi}=89.5$~{cm$^{-1}$}. This number is listed in parentheses in the first line of Tab.~\ref{tab_ljg_gra}.

Using the above fitted LJ potential, we can also calculate the interlayer frictional energy between two graphene layers. Fig.~\ref{fig_lj_cmode} shows the frictional energy for the relative shearing of two graphene layers along the x and y directions. Similar as the cohesive energy, the frictional energy is also in close relation with the interlayer phonon modes in bilayer graphene. The frictional energy curve around the minimum point determines the frequency of the C mode, which is shown in the right top inset in both panels of Fig.~\ref{fig_lj_cmode}. The frictional energy in the x-direction around the minimum point can be fit to the quadratic function $E=(1/2)\mu\omega^2(dx)^2$, in which $\omega$ is the frequency of the C$_x$ mode (with vibration along the x-direction). Similarly, the frictional energy in the y-direction gives the frequency of the C$_y$ mode (with vibration along the y-direction). For bilayer graphene, we find that $\omega_{Cx}=\omega_{Cy}=7.3$~{cm$^{-1}$}, which is smaller than the experimental value\cite{MohrM2007prb} by a factor of $\frac{1}{5}$. It implies that the frictional energy will be underestimated by a factor of $\frac{1}{25}$.

We have learned that the LJ potential is able to accurately describe the interlayer spacing and the cohesive energy between graphene layers. However, this potential has a friction issue; i.e., it underestimates the interlayer frictional energy in the bilayer graphene by one order. This is actually quite reasonable. Considering that there are only two parameters in the LJ potential, the prediction of this potential should be limited to two independent quantities only, i.e., the cohesive energy (B mode) and the interlayer spacing. We should not expect a good prediction for the third quantity of interest, the frictional energy (C mode). A straightforward solution for this friction issue is to increase the number of parameters in the potential model. For instance, seven additional parameters are introduced in Ref.~\onlinecite{KolmogorovAN2005} to resolve the friction issue.

\begin{figure}[tb]
  \begin{center}
    \scalebox{1.0}[1.0]{\includegraphics[width=8.0cm]{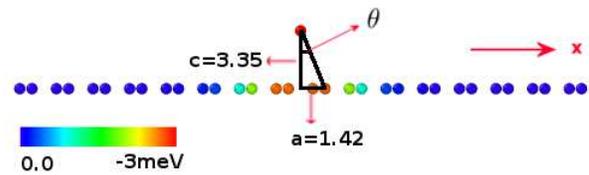}}
  \end{center}
  \caption{(Color online) Interlayer LJ potential for a carbon atom on top of a single layer of graphene. The six interlayer first-nearest-neighbor van der Walls bonds make most important contribution to the interlayer energy. The color bar is for the interlayer potential.}
  \label{fig_lj}
\end{figure}

\begin{figure}[tb]
  \begin{center}
    \scalebox{1.0}[1.0]{\includegraphics[width=8.0cm]{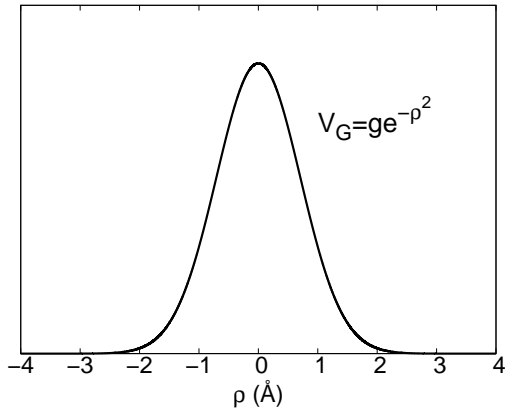}}
  \end{center}
  \caption{Gaussian shaped potential. $V_{G}=ge^{-\rho^2}$, with $\rho=\sqrt{x^2+y^2}$ as the projection of the distance onto the xy-plane. $g$ is the height of the potential.}
  \label{fig_gaussian}
\end{figure}

Before presenting our approach, we first make an explicit examination of this friction issue in the LJ potential. Fig.~\ref{fig_lj} shows that for a particular carbon atom from the top graphene layer, the LJ potential for this atom is mainly contributed by its six interlayer first-nearest-neighbor atoms in the bottom layer. We introduce an angle $\theta$ to describe the direction of these interlayer van der Waals bonds as shown in Fig.~\ref{fig_lj}. For bilayer graphene, we have $\tan\theta=b/c\approx0.42$, with $b=1.42$~{\AA} as the chemical C-C bond length in the graphene plane and $c=3.35$~{\AA} as the interlayer spacing. We get $\cos\theta=0.92$ and $\sin\theta=0.15$. For cohesive motion, the relative displacement between two graphene layers is $\vec{u}=u\hat{e}_z$. The resulting variation in the distance is $dr=u\hat{e}_z\cdot\hat{e}_r=u\cos\theta=0.92u$, where $\hat{e}_r=\vec{r}/r$ is the unit vector between two interacting atoms. For frictional motion (eg. in the x-direction), the relative displacement between two graphene layers is $\vec{u}=u\hat{e}_x$, so the resulting distance variation is $dr=u\sin\theta=0.15u$. This is much smaller than the distance variation induced by the cohesive motion. It indicates that frictional motion (in the x and y-directions) results in very small variations in the LJ potential, which is the underlying mechanism for the friction issue in the interlayer LJ potential for layered materials.

We find that the above friction issue for the LJ potential in layered materials can be eliminated by introducing only one more energy parameter. We propose to supplement the LJ potential by the following Gaussian shaped potential,
\begin{eqnarray}
V_{G} = g e^{-\rho^{2}},
\label{eq_g}
\end{eqnarray}
where $\rho=\sqrt{x^2+y^2}$ is the projection of the distance onto the xy-plane, and where $g$ is the only parameter for the Gaussian potential. The physical essence of this Gaussian potential is to guarantee the AA-stacking graphene layers to be the highest-energy configuration. This Gaussian potential impacts the frictional energy, but has no effect on the interlayer cohesive energy in the layered materials. Fig.~\ref{fig_gaussian} shows the Gaussian potential curve. The strength of the relative frictional motion is directly related to the frequency of the C mode, so the parameter $g=94.87$~{meV} is determined by fitting to the frequency of the C mode in bilayer graphene.

The total interaction energy between graphene layers is a combination of the LJ potential and the Gaussian potential,
\begin{eqnarray}
V = V_{LJ}+V_{G},
\label{eq_ljg}
\end{eqnarray}
where the LJ portion is calculated in Eq.~(\ref{eq_lj}) and the Gaussian portion is calculated in Eq.~(\ref{eq_g}). There are in total three potential parameters, with $\epsilon$ and $\sigma$ in the LJ potential and $g$ in the Gaussian potential.

\begin{table}
\caption{LJG parameters for bilayer MoS$_2$. Numbers in the parentheses are the experimental lattice constant\cite{WakabayashiN} and frequency.\cite{ZhangXprb2013} The first row is for the LJ potential ($g=0$), while the second row is the LJ-G potential. Energy parameter is in meV. Length parameter is in \AA.  Frequency is in cm$^{-1}$.}
\label{tab_ljg_mos2}
\begin{tabular}{|c|c|c|c|c|c|}
\hline 
$\epsilon$  & $\sigma$  & $g$   & $c$ (12.3)   & $\omega_{B}$ (40.2) & $\omega_{C}$ (23.1) \tabularnewline
\hline 
\hline 
23.6 & 3.18 & 0 & 12.36 & 40.2 & 14.5\tabularnewline
\hline 
23.6 & 3.18 & 175.68 & 12.36 & 40.2 & 23.1\tabularnewline
\hline 
\end{tabular}
\end{table}

\begin{table}
\caption{LJG parameters for bilayer BP. Numbers in the parentheses are experimental lattice constant\cite{BrownA1965ac} and frequency.\cite{YamadaY1984prb}  The first row is for the LJ potential ($g=0$), while the second row is the LJ-G potential. Energy parameter is in meV. Length parameter is in \AA. Frequency is in cm$^{-1}$.}
\label{tab_ljg_bp}
\begin{tabular}{|c|c|c|c|c|c|c|}
\hline 
$\epsilon$  & $\sigma$  & $g$ (meV) & $c$ (10.478)  & $\omega_{B}$ (61.6) & $\omega_{Cx}$ (13.7) & $\omega_{Cy}$ (36.5) \tabularnewline
\hline 
\hline 
15.94 & 3.438 & 0 & 10.5254 & 59.3 & 16.15 & 18.11\tabularnewline
\hline 
15.94 & 3.438 & 123.0 & 10.5254 & 59.3 & 29.0 & 36.2\tabularnewline
\hline 
\end{tabular}
\end{table}

Following the same procedure as bilayer graphene, we can obtain LJ-G potential for MoS$_2$ bilayers and BP bilayers. Tab.~\ref{tab_ljg_mos2} lists LJ-G potential parameters for the MoS$_2$ layers, while Tab.~\ref{tab_ljg_bp} shows the LJ-G potential parameters for the BP layers. The fitting procedure to obtain the three parameters $\sigma$, $\epsilon$ and $g$ is the same as for graphene, i.e. they were obtained by fitting to the out-of-plane lattice constant, interlayer breathing mode, and interlayer shear mode, respectively. It should be noted that BP is highly anisotropic in the two in-plane directions resulting from its puckered configuration, which leads to different frequencies for the two C modes in bilayer BP.\cite{JiangJW2014BCmode} The LJ-G potential can only provide an accurate description for one C mode, since there is only one potential parameter in the Gaussian potential. Potentials with at least two parameters (eg. two independent Gaussian potentials) are needed to describe accurately both C modes in BP layers. 

One advantage of the LJ-G potential proposed here is that the potential parameters for heterostructures constructed using different layered materials can be extracted using the standard geometric combination rules
\begin{eqnarray}
\epsilon & = & \sqrt{\epsilon_{1}\epsilon_{2}}\nonumber\\
\sigma & = & \frac{\sigma_{1}+\sigma_{2}}{2}\\\nonumber
A & = & \sqrt{A_{1}A_{2}}.
\label{eq_comb_rule}
\end{eqnarray}
Hence, the LJ-G potential can be easily applied to study the interlayer interactions in graphene/MoS$_2$/BP heterostructures.

\section{Cohesive and Frictional Energy}

\begin{figure}[tb]
  \begin{center}
    \scalebox{1.0}[1.0]{\includegraphics[width=8.0cm]{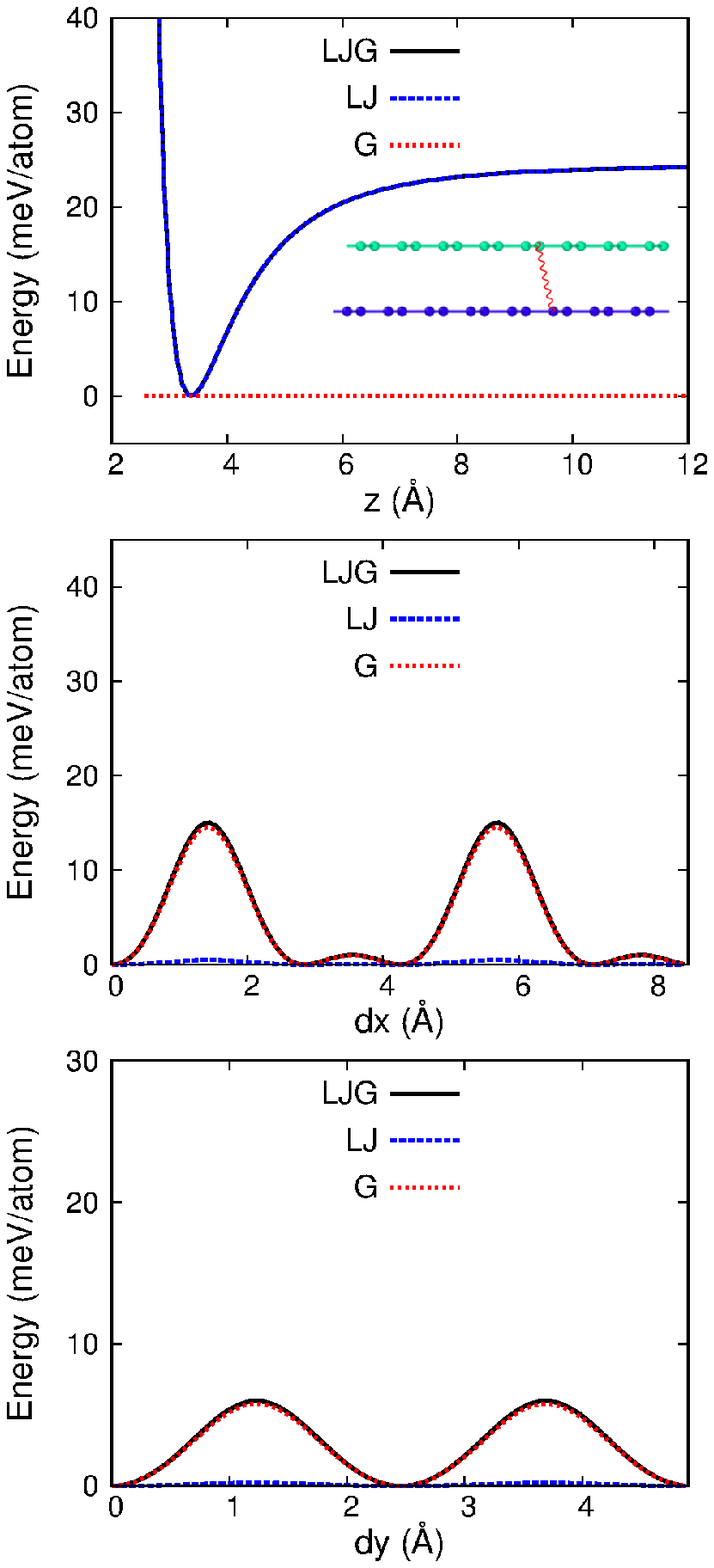}}
  \end{center}
  \caption{(Color online) Energy variations for bilayer graphene from the LJ and Gaussian potentials. Top panel: Gaussian potential has no contribution to the cohesive energy. Middle and bottom panels: Gaussian potential contributes 96.1\% of the frictional energy along x and y directions. The spring in the inset indicates the interlayer interaction.}
  \label{fig_energy_graphene}
\end{figure}

\begin{figure}[tb]
  \begin{center}
    \scalebox{1.0}[1.0]{\includegraphics[width=8.0cm]{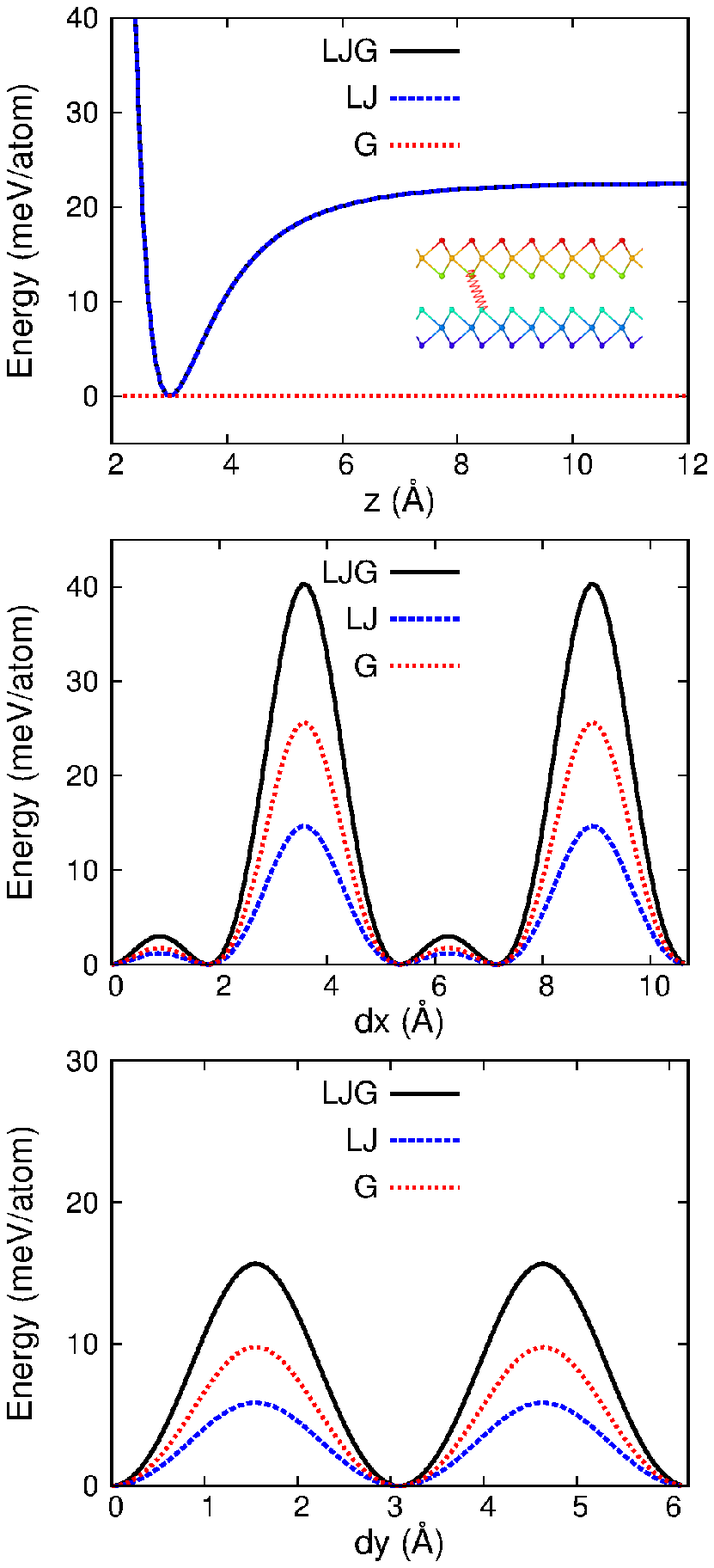}}
  \end{center}
  \caption{(Color online) Energy variations for bilayer MoS$_2$ from LJ potential and Gaussian potential. Top panel: Gaussian potential has no contribution to the cohesive energy. Middle and bottom panels: Gaussian potential contributes 60.3\% of the frictional energy along x and y directions. The spring in the inset indicates the interlayer interaction.}
  \label{fig_energy_mos2}
\end{figure}

\begin{figure}[tb]
  \begin{center}
    \scalebox{1.0}[1.0]{\includegraphics[width=8.0cm]{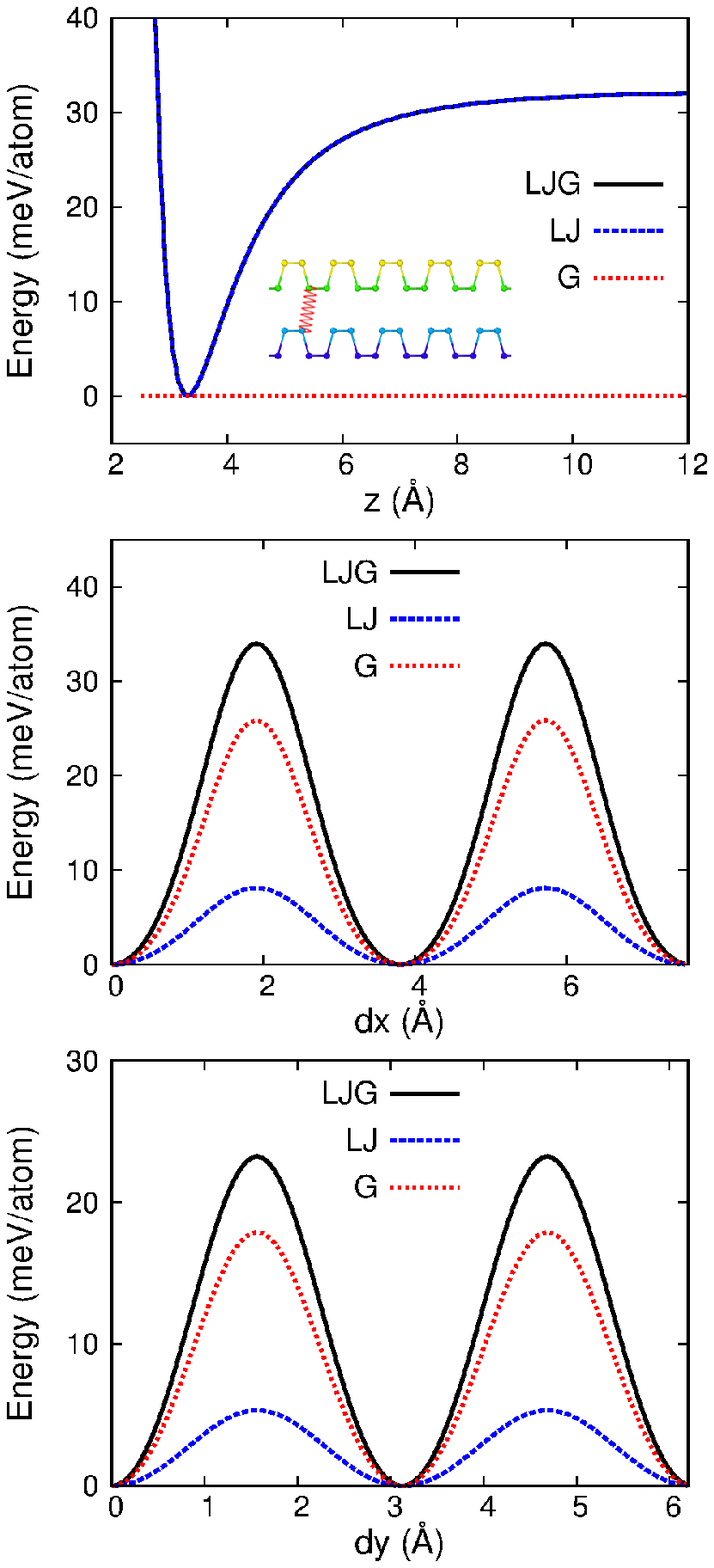}}
  \end{center}
  \caption{(Color online) Energy variations for bilayer BP from LJ potential and Gaussian potential. Top panel: Gaussian potential has no contribution to the cohesive energy. Middle and bottom panels: Gaussian potential contributes 75.0\% of the frictional energy along x and y directions. The spring in the inset indicates the interlayer interaction.}
  \label{fig_energy_bp}
\end{figure}

\begin{figure}[tb]
  \begin{center}
    \scalebox{1.0}[1.0]{\includegraphics[width=8.0cm]{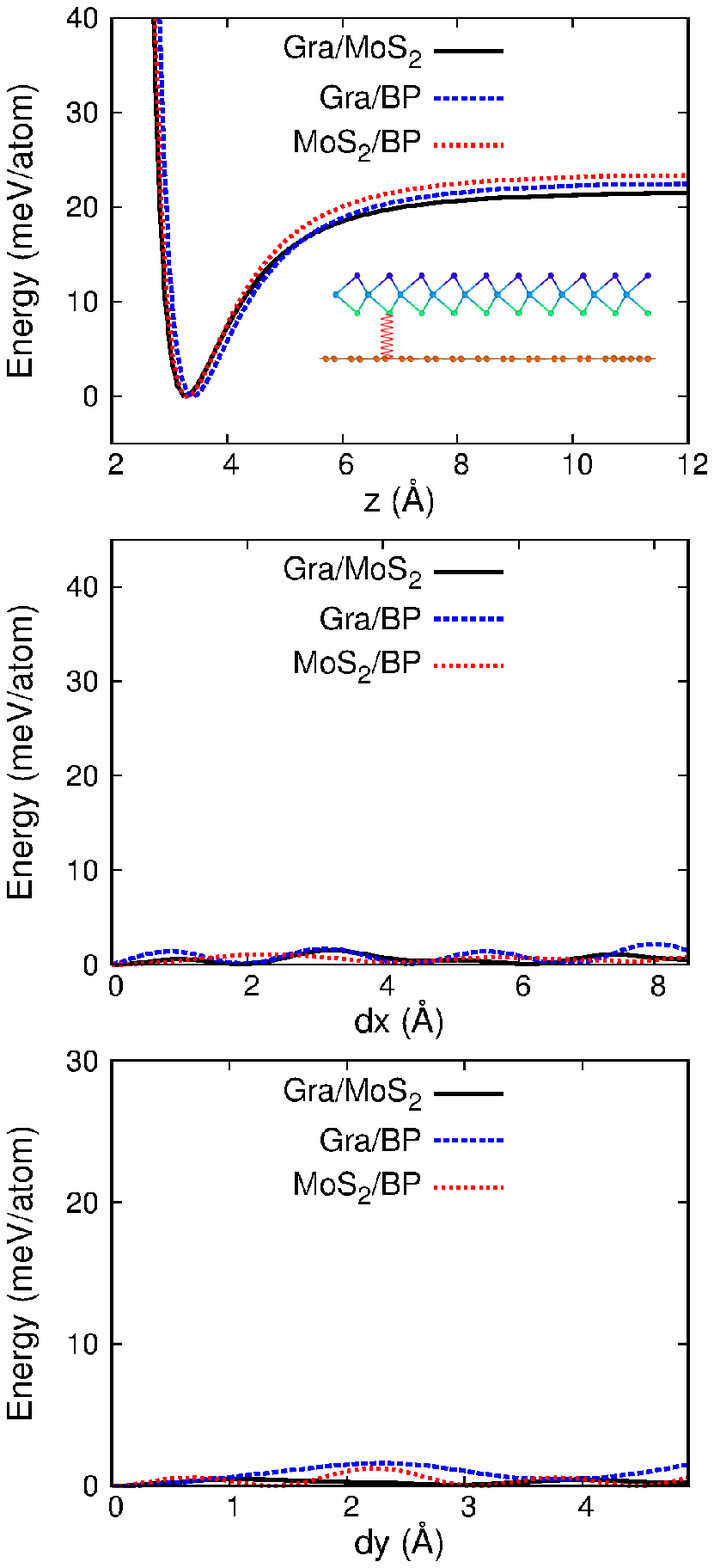}}
  \end{center}
  \caption{(Color online) Energy variations for graphene/MoS$_2$, graphene/BP, and MoS$_2$/BP heterostructures. The interlayer interaction is described by the LJ-G potential. Top panel: cohesive energies for the heterostructures are almost the same as the individual compositions. Middle and bottom panels: frictional energies for the heterostructures are one order smaller than the individual constituents. The spring in the inset indicates the interlayer interaction.}
  \label{fig_energy_heterostructure}
\end{figure}

In the previous section, we have proposed the LJ-G potential to describe the interlayer interaction of layered materials. The rest of this paper is devoted to the application of the LJ-G potential for computing the cohesive energy and frictional energy in graphene, MoS$_2$, BP, and their heterostructures.

Fig.~\ref{fig_energy_graphene} top panel shows the cohesive energy in bilayer graphene. The zero energy point is set at the equilibrium interlayer spacing for bilayer graphene. The cohesive energy is contributed solely by the LJ potential, while the Gaussian potential has no contribution. The cohesive energy in the limit of large interlayer spacing is 24.3~{meV/atom}, which can be regarded as the cohesion energy for bilayer graphene. This cohesive energy value is in good agreement with the value of 23.0~{meV/atom} from the first principles calculations.\cite{GirifalcoLA2002prb} The frictional energy curves along the x and y directions are shown in the middle and bottom panels. It can be seen that the LJ potential makes a very limited contribution to the total frictional energy, which is dominated by the Gaussian potential. From the middle panel, the AA-stacking bilayer graphene has the maximum frictional energy of 15.0~{meV/atom}, which is in the range of previously reported values of 13.0~{meV/atom} in Ref~\onlinecite{KolmogorovAN2005} and 19.0~{meV/atom} in Ref~\onlinecite{LebedevaIV}.

Fig.~\ref{fig_energy_mos2} shows the cohesive energy and the frictional energy in bilayer MoS$_2$. The cohesive energy for bilayer MoS$_2$ is 22.6~{meV/atom}, which is quite close to the cohesive energy of bilayer graphene. For the frictional energy, an obvious difference between bilayer MoS$_2$ and bilayer graphene is that the contribution from the LJ potential for the frictional energy in bilayer MoS$_2$ is 39.7\%, which is considerably larger than 3.9\% in bilayer graphene. The contribution percentage of each individual potential is computed based on the energy variation around the energy minimum point.  Furthermore, the maximum frictional energy for bilayer MoS$_{2}$ is significantly larger than bilayer graphene, reaching about 40~{meV/atom} in the x-direction and about 15~{meV/atom} in the y-direction.

The cohesive energy and frictional energy for the bilayer BP is shown in Fig.~\ref{fig_energy_bp}. The cohesive energy is 32.1~{meV/atom} which is larger than both bilayer graphene and MoS$_2$. The LJ potential contributes 25\% to the frictional energy in bilayer BP.  The maximum frictional energy for bilayer BP is in between that of bilayer graphene and bilayer MoS$_{2}$, reaching about 33~{meV/atom} in the x-direction and about 22~{meV/atom} in the y-direction.

For the graphene/MoS$_2$, graphene/BP, and MoS$_2$/BP heterostructures, the LJ-G potential parameters are determined by the combination rule in Eq~(\ref{eq_comb_rule}). Fig.~\ref{fig_energy_heterostructure} top panel shows that the cohesive energy of all three heterostructures are very similar. The frictional energy for these heterostructures are at least one order smaller than each individual constituent. The weak frictional energy in the heterostructure is due to the intrinsic lattice mismatch of the two individual constituents.\cite{WangLF2014nano} This weak frictional energy can also be analyzed from a geometrical point of view.\cite{JiangJW2012japmoire} The intrinsic lattice mismatch leads to a Moir{\'e} pattern, resulting in a large unit cell for the heterostructure. The Moir{\'e} pattern varies during the frictional motion of the heterostructure. The large unit cell contains lots of inequivalent atoms; i.e., these atoms have different contribution to the interlayer interaction. The total interlayer energy is the summation over the potential for all of these inequivalent atoms. Mathematically, the summation can be regarded as an integration, which is independent of the details for the Moir{\'e} pattern. Hence, the interlayer potential remains almost unchanged during the frictional motion.

\section{conclusion}
In conclusion, we have proposed a Gaussian potential with only one parameter to supplement the standard LJ potential in layered materials such as graphene, MoS$_2$, BP, and their heterostructures. The Gaussian potential governs the frictional motion of the layered system, while it has no effect on the cohesive motion for layered materials.  The LJ-G potential energy parameters are fitted to the frequency of the interlayer B mode and C mode. As an application of the LJ-G potential, we calculated the interlayer cohesive energy and frictional energy in graphene, MoS$_2$, BP, and their heterostructures. Due to the intrinsic lattice mismatch in the heterostructure, the frictional energy for the heterostructure is found to be one order smaller than the frictional energy in each individual constituent.  

\textbf{Acknowledgements} The work is supported by the Recruitment Program of Global Youth Experts of China and the start-up funding from Shanghai University. HSP acknowledges the support of the Mechanical Engineering department at Boston University.

%
\end{document}